\documentclass[10pt,conference]{IEEEtran}
\IEEEoverridecommandlockouts
\usepackage{cite}
\usepackage{amsmath,amssymb,amsfonts}
\usepackage{algorithmic}
\usepackage{url}
\usepackage{graphicx}
\usepackage{textcomp}
\usepackage[caption=false,font=footnotesize]{subfig}
\usepackage[table]{xcolor}
\usepackage{booktabs} 
\usepackage{threeparttable}

\usepackage{makecell} 
\usepackage{multirow} 
\def\BibTeX{{\rm B\kern-.05em{\sc i\kern-.025em b}\kern-.08em
    T\kern-.1667em\lower.7ex\hbox{E}\kern-.125emX}}
\begin{document}

\title{Beyond Paper-to-Paper: Structured Profiling and Rubric Scoring for Paper-Reviewer Matching
\thanks{This work was supported in part by
the Natural Science Foundation of China under Grant No. T2322027 and
62442204, and the Key Research Program of the Chinese Academy
of Sciences under Grant No. RCJJ-145-24-20.}
}

\author{
\IEEEauthorblockN{Yicheng Pan\textsuperscript{1,$\dagger$}, Zhiyuan Ning\textsuperscript{2,$\dagger$}, Ludi Wang\textsuperscript{3}, Yi Du\textsuperscript{1,3,*}}
\IEEEauthorblockA{
\textsuperscript{1}Hangzhou Institute for Advanced Study, University of Chinese Academy of Sciences, Hangzhou, China \\
\textsuperscript{2}Westlake University, Hangzhou, China \\
\textsuperscript{3}Computer Network Information Center, Chinese Academy of Sciences, Beijing, China \\
panyicheng25@mails.ucas.ac.cn, ningzhiyuan@westlake.edu.cn, \{wld, duyi\}@cnic.cn \\
\textsuperscript{$\dagger$}These authors contributed equally.  
\textsuperscript{*}Corresponding author.
}
}
\maketitle
\begin{abstract}
As conference submission volumes continue to grow, accurately recommending suitable reviewers has become a  challenge. Most existing methods follow a ``Paper-to-Paper'' matching paradigm, implicitly representing a reviewer by their publication history. However, effective reviewer matching requires capturing multi-dimensional expertise, and textual similarity to past papers alone is often insufficient. To address this gap, we propose P2R, a training-free framework that shifts from implicit paper-to-paper matching to explicit profile-based matching. P2R uses general-purpose LLMs to construct structured profiles for both submissions and reviewers, disentangling them into Topics, Methodologies, and Applications. Building on these profiles, P2R adopts a coarse-to-fine pipeline to balance efficiency and depth. It first performs hybrid retrieval that combines semantic and aspect-level signals to form a high-recall candidate pool, and then applies an LLM-based committee to evaluate candidates under strict rubrics, integrating both multi-dimensional expert views and a holistic Area Chair perspective. Experiments on NeurIPS, SIGIR, and SciRepEval show that P2R consistently outperforms state-of-the-art baselines. Ablation studies further verify the necessity of each component. Overall, P2R highlights the value of explicit, structured expertise modeling and offers practical guidance for applying LLMs to reviewer matching.
\end{abstract}


\section{Introduction}
Peer review is central to trustworthy science, yet rising submission volumes~\cite{weijun2023identifying,xiao2025interdisciplinary,qiao2022rpt,cai2023resolving,ma2023comprehensive} make fully manual paper--reviewer assignment increasingly impractical. This motivates \emph{Automatic Paper--Reviewer Matching}: given a submission, the goal is to identify the most suitable and competent experts to ensure efficient and high-quality reviewing~\cite{price2017computational,wang2026ai}. The key question is how to represent reviewer expertise in a way that reflects the actual decision criteria used by Area Chairs, while remaining scalable to large conferences.

Prior work has explored a range of strategies. Early methods typically concatenate a reviewer's historical publications into a single document and compute keyword similarity (e.g., TF-IDF) against the submission~\cite{liu2014robust}. Graph-based methods~\cite{ning2025rethinking,qiao2020context,ning2025deep,dong2023adaptive,ning2021lightcake,ning2022graph} construct networks from co-authorship or citations and propagate expertise scores through the graph structure~\cite{tong2006fast}. Embedding-based approaches~\cite{ning2024fedgcs} encode papers into dense vectors to measure semantic proximity~\cite{beltagy2019scibert,cohan2020specter,singh2023scirepeval}. Most recently, Chain-of-Factors (CoF) learns factor-aware representations to retrieve similar papers and aggregates paper-level scores to rank reviewers~\cite{zhang2025chain}. Despite their differences, these approaches share a common core: they model relationships between papers and implicitly  represent a reviewer as a composite of their publications.

\begin{figure}[t] 
    \centering
    \includegraphics[width=\columnwidth]{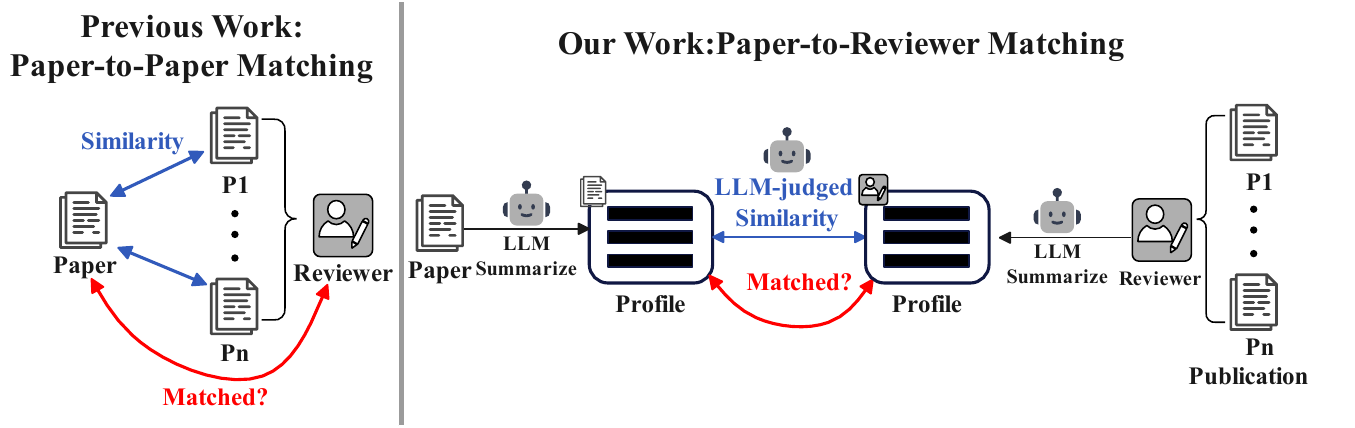} 
    
\caption{\textbf{Paradigm comparison.} The left panel illustrates traditional \textbf{Paper-to-Paper} matching, which relies on similarities between a submission and historical publications. In contrast, the right panel depicts our \textbf{Paper-to-Reviewer} framework (P2R). P2R leverages LLMs to synthesize structured profiles, enabling direct and interpretable expertise alignment.}
    \label{fig:comparison}
\end{figure}

However, reviewer assignment is a fundamentally multi-factor decision. In practice, Area Chairs must assess the specific \emph{paper--reviewer fit}, which depends on explicit dimensions such as the research topics, the methodological toolkit, and the target application setting~\cite{singh2023scirepeval}. A simple paper-to-paper match often misses these critical differences. For example, a reviewer may publish extensively on the same topics as a submission but primarily conduct theoretical analysis rather than the empirical evaluation required by the paper, or work in a different application domain. Such structural mismatches are difficult to resolve by aggregating paper similarities. This motivates moving from paper-based proxies toward a multi-dimensional characterization of both submissions and reviewers.

Recent Large Language Models (LLMs) have demonstrated strong capabilities in long-context understanding, instruction following, and reasoning over scientific text. Leveraging these capabilities, we propose \textbf{P2R}, a multi-stage framework that shifts from implicit paper-to-paper matching to explicit profile-based matching. P2R constructs structured profiles for submissions and reviewers and then performs a committee-style assessment under explicit rubrics. Concretely, P2R consists of three stages: (1) \textbf{Structured Profiling}, which generates profiles along three dimensions---Topics, Methodologies, and Applications; (2) \textbf{Hybrid Retrieval}, which combines profile matching with embedding retrieval to efficiently shortlist high-recall candidates; and (3) \textbf{Rubric-driven Committee Scoring}, where an LLM committee consisting of three dimension-specific evaluators and a holistic Area Chair scores candidates under explicit rubrics to produce the final ranking.

We validate the effectiveness of P2R through  experiments on three benchmarks: NeurIPS, SIGIR, and SciRepEval. P2R achieves the best overall performance across all three datasets, surpassing strong pretrained language model baselines and CoF in a \emph{training-free} manner, in contrast to approaches that rely on task-specific training~\cite{singh2023scirepeval,zhang2025chain}. Comprehensive ablation studies further show that each stage is indispensable and contributes complementary gains, supporting the value of explicit profiling and rubric-based evaluation.

Overall, P2R advances reviewer matching by directly modeling paper--reviewer fit with  multi-dimensional signals rather than relying solely on paper representations. Beyond improved accuracy, our framework offers a practical recipe for applying LLMs to reviewer matching: use LLMs to build structured expertise profiles, retrieve candidates with high recall, and enforce decision criteria via rubrics during final scoring.

To support reproducibility, the source code is publicly available at: \url{https://github.com/kg4sci/P2R}.

\section{RELATED WORK}

\subsection{Automatic Paper--Reviewer Matching}
Automatic paper--reviewer matching has been widely studied to reduce the manual burden in large-scale peer review.
Most pipelines separate (i) learning a relevance function $f(p,r)$ between a submission $p$ and a reviewer $r$, and (ii) solving a constrained assignment problem with capacity and conflict-of-interest constraints\cite{charlin2011framework,schrijver2003combinatorial}. 
We focus on the first part and aim to build a relevance function that is both accurate and interpretable.
Early systems rely on lexical similarity and topic models, while graph-based methods further exploit co-authorship and citation links to propagate expertise signals\cite{tong2006fast,saveski2023counterfactual}. 
 Production systems such as TPMS adopt human-in-the-loop workflows: they provide ranked lists to Area Chairs and use bids to refine relevance estimates\cite{charlin2013toronto}.   
These approaches are strong baselines, but expertise is often reduced to coarse textual or topical similarity\cite{mimno2007expertise}.  Some work further formulates reviewer assignment as constraint-based optimization and studies robustness under sparse or noisy evidence\cite{tang2010expertise,liu2014robust}.

\subsection{Scientific Document Representations for Matching}
Recent work formulates matching as semantic retrieval. Domain PLMs such as SciBERT provide scientific representations, and citation-informed models such as SPECTER and SciNCL further improve embeddings with citation graphs\cite{beltagy2019scibert,cohan2020specter,ostendorff2022neighborhood}. 
They are commonly evaluated on benchmarks such as SciRepEval\cite{singh2023scirepeval}. 
To address the limits of single-vector matching, Chain-of-Factors (CoF) combines semantic, topics, and citation signals and applies instruction tuning in a coarse-to-fine framework\cite{zhang2025chain}. 
However, it still depends on implicit embedding similarity, which can blur distinct expertise dimensions and hide structural mismatches.
We instead use explicit multi-dimensional profiles to make mismatches visible.

\subsection{LLMs for Reviewer Recommendation and Decision Making}
LLMs increasingly drive reviewer recommendation\cite{hou2024large,karan2025dataset,peng2025frontier}. 
Recent benchmarks use LLMs as zero-shot rankers with pairwise prompting, showing strong direct preference judgments\cite{sun2023chatgpt}. 
LLMs also perform well as rubric-based judges, motivating committee-style deliberation rather than a single similarity score\cite{gu2024survey,zheng2023judging}. 
This is supported by the emerging capabilities of LLMs in complex reasoning and multi-step decision making, often elicited through advanced prompting strategies.
We follow this direction but enforce explicit structure.
We replace generic summaries with schema-consistent profiling that separates \emph{Topics}, \emph{Methodologies}, and \emph{Applications}.
We then apply rubric scoring with multiple expert perspectives, which moves from latent correlation to explicit decision simulation.

\section{PROPOSED METHOD}
\begin{figure*}[htbp]
    \centering
    \includegraphics[width=1\textwidth]
    {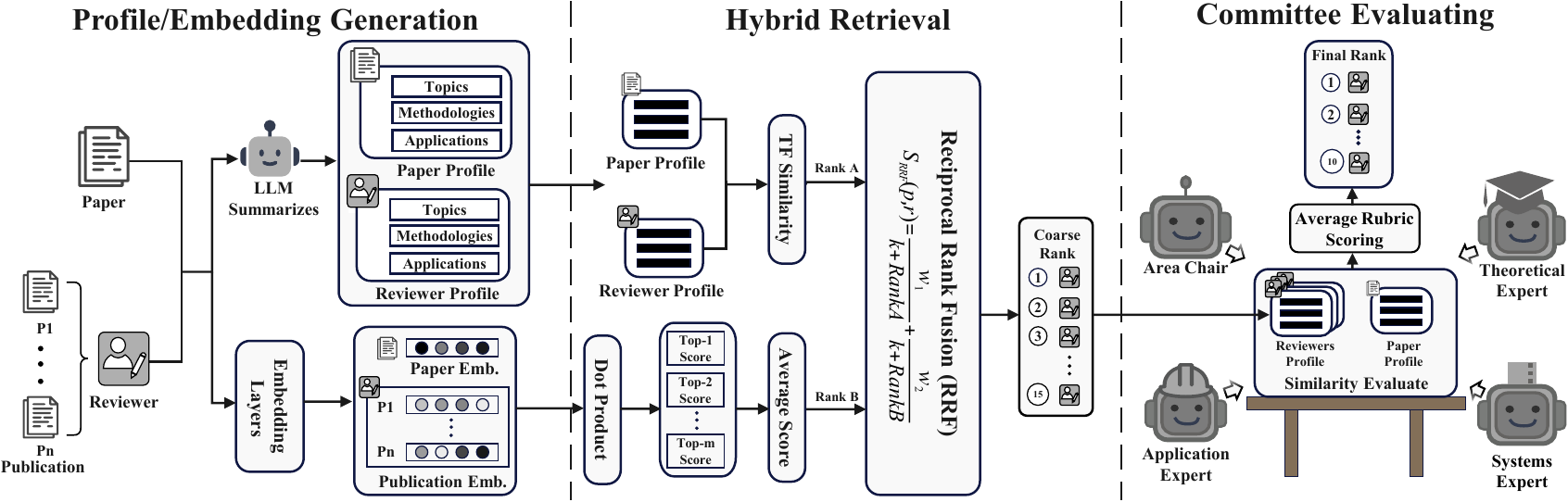} 
\caption{Overview of P2R. The pipeline has three stages: (1) Profile and embedding generation, where an LLM summarizes structured aspect profiles for both the submission and each reviewer (from their historical publications), and an embedding model encodes the submission and all reviewer-history papers into dense vectors; (2) Hybrid retrieval, which fuses discrete profile matching and semantic matching via RRF to obtain a coarse ranking; and (3) Committee evaluation, where an LLM committee scores candidates with a rubric from multiple expert perspectives to produce the final ranking.}

    \label{fig:framework_large}
\end{figure*}
\subsection{Problem Setup}

Given a submission paper $p$ with text $x_p$ (title and abstract) and a reviewer $r$ with $n_r$ historical publications,
let $x_{r,i}$ denote the concatenation of the title and abstract of the $i$-th historical paper.
We further define the reviewer digest as $\bar{x}_r=\mathrm{concat}(x_{r,1},\ldots,x_{r,n_r})$.
The goal is to produce a ranked list of reviewers for each submission.
In this work, we focus on designing a relevance scoring function $f(p,r)$ for ranking,
while leaving conflict constraints to downstream assignment.

\subsection{Granularity-aware Expertise Profiling}
Dense embeddings often collapse distinct competence dimensions into a single scalar, obscuring structural mismatches (e.g., a shared topic but differing methodologies). To explicitly represent expertise, we leverage the instruction-following capabilities of LLMs to construct explicit, multi-dimensional reviewer profiles. We define a structured profile schema $\mathcal{D} = \{\textsc{Topics}, \textsc{Methodologies}, \textsc{Applications}\}$, corresponding to the critical dimensions of paper-reviewer fit.

For any scientific text $x$, we employ a general-purpose LLM (Qwen3) \cite{yang2025qwen3} as a profiler  to extract a structured profile $\phi(x)$:
\begin{equation}
\label{eq:profile_struct}
\phi(x) = \big\{ \mathcal{L}_d(x) \mid d \in \mathcal{D} \big\},
\end{equation}
where $\mathcal{L}_d(x)$ represents a list of concise phrases for dimension $d$ (e.g., specific algorithms for \textsc{Methodologies}).
\noindent\textbf{Reviewer profiling.}
For each reviewer $r$, we query an instruction-following LLM (Qwen3) on the reviewer digest $\bar{x}_r$ with a structured instruction, e.g., \textit{``Analyze the following publication history of a NeurIPS reviewer. Extract the following fields into a JSON object: 1. `topics' (3-5 high-level research topics); 2. `methodologies' (3-5 algorithmic approaches); 3. `applications' (3-5 specific domains)...''}.
This prompts the model to output a JSON-only profile $\phi(\bar{x}_r)$ under a strict schema constraint. If a field cannot be supported by evidence, the profiler returns an empty list, reducing hallucination risk and ensuring schema consistency.

\noindent\textbf{Submission profiling.}
Similarly, for each submission $p$, we apply the same profiler to the title and abstract to obtain $\phi(x_p)$. This creates a shared structured representation space between papers and reviewers, enabling explicit aspect-level matching beyond latent semantic similarity.

\noindent\textbf{Implementation details.}
P2R is training-free: it does not finetune any model on target datasets. Profiling is performed via Qwen3 API with JSON-only prompting; we use a low temperature and repeat each run three times to mitigate stochasticity, reporting mean performance.

\subsection{Hybrid Retrieval with Reciprocal Rank Fusion}
P2R performs coarse candidate selection by fusing two complementary signals: (i) discrete matching over structured profiles, and (ii) continuous similarity via  embeddings. Let $\mathcal{R}$ denote the set of candidate reviewers.

\noindent\textbf{Stream A: Discrete profile matching.}
To enable precise lexical matching, we linearize the structured profile into a discrete string $z(x)$ by joining all phrases across dimensions $\mathcal{D}$ with a space separator:
\begin{equation}
\label{eq:flatten}
z(x)=\mathrm{concat}\Big(\big(\mathrm{concat}(\mathcal{L}_d(x))\big)_{d\in\mathcal{D}}\Big),
\end{equation}
where $\mathrm{concat}(\cdot)$ denotes string concatenation and dimensions are concatenated in a fixed order: \textsc{Topics}, \textsc{Methodologies}, then \textsc{Applications}.

We thus obtain the linearized submission profile $z_p = z(x_p)$ and reviewer profile $z_r = z(\bar{x}_r)$.
We compute TF-cosine similarity:
\begin{equation}
s_{\text{tf}}(p,r) = \cos\big(\mathrm{tf}(z_p), \mathrm{tf}(z_r)\big).
\end{equation}
$\mathrm{tf}(\cdot)$ denotes raw term-frequency vectors over the profile vocabulary, $\ell_2$-normalized before cosine similarity.
This stream emphasizes exact overlaps of explicit expertise facets (e.g., application domains or method names), which are often diluted in dense embeddings.

\noindent\textbf{Stream B: General-purpose LLM embeddings over reviewer history.}
To obtain a robust semantic signal without relying on scientific-domain Pre-trained Language Model, we use NV-Embed as a general-purpose dense embedding model\cite{lee2024nv}. 
NV-Embed is a decoder-only LLM-based embedder designed for retrieval and broad text embedding tasks, which employs a dedicated pooling mechanism (latent attention) and contrastive training recipes to produce strong sentence representations.
In our framework, we do not perform any additional instruction tuning or finetuning; we directly use the released NV-Embed checkpoint as an off-the-shelf encoder.

Given a submission paper $p$, we compute its  embedding $e_p$ from its title and abstract. 
Similarly, for each reviewer $r$, we compute embeddings $\{e_{r,i}\}$ for all their historical papers.
We define the reviewer--paper semantic score as the mean cosine similarity over the Top-$m$ most similar history papers:
\begin{equation}
s_{\text{emb}}(p,r)=\frac{1}{m}\sum_{i\in\mathrm{Top}\text{-}m}\big(e_p^\top e_{r,i}\big),
\end{equation}
where all vectors are $\ell_2$-normalized and, to be consistent with prior work, we set $m=3$.
This Top-$m$ aggregation effectively reduces noise from irrelevant historical publications.

\noindent\textbf{Rank Fusion and Candidate Generation.}
To combine these heterogeneous signals, we convert scores to rankings.
Let $\mathrm{RankA}$ and $\mathrm{RankB}$ denote the rank of reviewer $r$ for submission $p$ derived from $s_{\text{tf}}$ and $s_{\text{emb}}$, respectively.
 We compute the coarse retrieval score via Reciprocal Rank Fusion (RRF)\cite{cormack2009reciprocal}:
\begin{equation}
s_{\text{rrf}}(p,r)=\frac{w_{1}}{k+\mathrm{RankA}} + \frac{w_{2}}{k+\mathrm{RankB}},
\end{equation}
where $k$ is a smoothing constant and $w_{\text{1}}, w_{\text{2}}$ are stream weights. 
We retain the Top-$M$ reviewers as the candidate set $\mathcal{C}_p$ for the subsequent committee evaluation stage. 
We set $M=15$ to ensure a sufficient candidate pool for the  committee evaluation.

\subsection{Rubric-driven Committee Evaluation}
Although hybrid retrieval effectively aggregates lexical and semantic signals to optimize recall, it relies on generalized similarity scores that may mask mismatches in specific dimensions. To better ensure expertise matching, we leverage LLMs to simulate a multi-perspective Area Chair committee. Instead of relying on a single generic score, we query the model under $C$ distinct personas (i.e., Area Chair, Theoretical Expert, Application Expert, and Systems Expert) to ensure a comprehensive evaluation.

\noindent\textbf{Aspect-based Rubric Scoring.}
Each  committee member $c$ evaluates the alignment between profiles $\phi(x_p)$ and $\phi(\bar{x}_r)$ against an explicit rubric, e.g., \textit{``You are a SIGIR Area Chair... Three-dimension rubric: Topics / Methodologies / Applications...
For each paper tag, find its best match among the  reviewer tags:
strong match $=1.0$ (near-synonymous), related match $=0.5$, otherwise $0$; sum these weights to obtain the per-dimension overlap $\mathrm{overlap}_d$...''}.
Let $\mathcal{L}_{p}^{d} = \mathcal{L}_d(x_p)$ and $\mathcal{L}_{r}^{d} =   \mathcal{L}_d(\bar{x}_r)$ denote the tag lists of the paper and reviewer in dimension $d$, respectively.
The LLM computes one-to-one overlap and assigns per-dimension scores using coverage and softJaccard thresholds
(e.g., where for a dimension $d$,
$\mathrm{coverage}_d=\mathrm{overlap}_d/|\mathcal{L}_{p}^{d}|$ and
$\mathrm{softJaccard}_d=\mathrm{overlap}_d/(|\mathcal{L}_{p}^{d}|+|\mathcal{L}_{r}^{d}|-\mathrm{overlap}_d)$; a high match is assigned (e.g., $s_d=3$) if $\mathrm{coverage}_d$ and $\mathrm{softJaccard}_d \ge 2/3$, and a lower $s_d$ otherwise).
We use $\mathrm{coverage}_d$ to ensure sufficient requirement coverage and $\mathrm{softJaccard}_d$ to control profile specificity, preventing high scores from reviewers with overly generic tag sets.
The final score is computed as a weighted sum of aspect scores:
\begin{equation}
s^{(c)}_{\text{cont}}(p,r)  = \alpha \cdot s_T + \beta \cdot s_M + \gamma \cdot s_A,
\end{equation}
where $s_T$, $s_M$, and $s_A$ denote the rubric scores for \textsc{Topics}, \textsc{Methodologies}, and \textsc{Applications}, respectively, and $\alpha, \beta, \gamma$ represent their corresponding importance coefficients.
To match the annotation schema, we map the continuous score $s^{(c)}_{\text{cont}}(p,r)$ into the discrete label space $\{0,1,2,3\}$ using three thresholds $\tau_1 < \tau_2 < \tau_3$, assigning labels $0$, $1$, $2$, and $3$ to the intervals $(-\infty,\tau_1)$, $[\tau_1,\tau_2)$, $[\tau_2,\tau_3)$, and $[\tau_3,\infty)$, respectively. This step is used only to align predictions with the ground truth label space.

\noindent\textbf{Committee aggregation.}
Let $s^{(c)}_{\text{llm}}(p,r)$ be the final discrete score assigned by the $c$-th committee member. We aggregate the judgments from all $C$ members by computing the mean:
\begin{equation}
s_{\text{llm}}(p,r)=\frac{1}{C}\sum_{c=1}^{C}s^{(c)}_{\text{llm}}(p,r).
\end{equation}

\noindent\textbf{Final fusion with retrieval.}
We define the final relevance function as $f(p,r)=s_{\text{final}}(p,r)$.
Let $\mathrm{Rank}_{\text{llm}}(p,r)$ denote the rank of reviewer $r$ for submission $p$ derived from the committee score $s_{\text{llm}}$.
We treat the committee ranking as an additional RRF stream with weight $w_3$ :
\begin{equation}
s_{\text{final}}(p,r)=s_{\text{rrf}}(p,r)+\frac{w_{3}}{k+\mathrm{Rank}_{\text{llm}}(p,r)}.
\end{equation}

This preserves high-recall retrieval signals while allowing rubric-based evaluation to correct misranked candidates.

\section{Experiments}
\subsection{Experiment Setup}
\noindent\textbf{Datasets.}
Following prior works~\cite{mimno2007expertise,karimzadehgan2008multi,singh2023scirepeval}, We evaluate on three public reviewer-matching benchmarks with expert-curated relevance labels: NeurIPS~\cite{mimno2007expertise}, SIGIR~\cite{karimzadehgan2008multi}, and SciRepEval~\cite{singh2023scirepeval}. Dataset statistics are summarized in Table~\ref{tab:stats}. NeurIPS uses a 0--3 relevance scale; SIGIR derives relevance from aspect-level overlap by discretizing the Jaccard similarity between paper and reviewer aspect labels into the same 0--3 scale; SciRepEval unifies augmented NeurIPS and ICIP ratings into the same 0--3 scale.
\noindent\textbf{Baselines.}
We compare P2R with the lexical matching baseline TPMS~\cite{charlin2013toronto}.
We include scientific PLM retrievers SciBERT~\cite{beltagy2019scibert}, SPECTER~\cite{cohan2020specter}, and SciNCL~\cite{ostendorff2022neighborhood}, as well as the general-purpose dense retriever COCO-DR~\cite{yu2022coco}.
We also evaluate SPECTER 2.0~\cite{singh2023scirepeval} using both PRX and CLF variants under its adapter-based multi-task setting.
Finally, we compare with the state-of-the-art Chain-of-Factors (CoF)~\cite{zhang2025chain}, which combines semantic, topic, and citation factors and applies a coarse-to-fine ranking pipeline.

\begin{table}[!t]
\centering
\caption{Dataset Statistics}
\label{tab:stats}
\setlength{\tabcolsep}{1.5pt} 
\renewcommand{\arraystretch}{1.2} 

\resizebox{\columnwidth}{!}{%
    \begin{tabular}{l||cccc}
    \hline
    Dataset & \#Papers & \#Reviewers & Conference(s) & \# Annotated Pairs \\
    \hline
    NeurIPS & 34 & 190 & NeurIPS 2006 & 393 \\
    SIGIR & 73 & 189 & SIGIR 2007 & 13797 \\
    SciRepEval & 107 & 661 & NeurIPS 2006, ICIP 2016 & 1729 \\
    \hline
    \end{tabular}%
}
\end{table}

\begin{table*}[!t]
    \centering
    \caption{Performance comparison with state-of-the-art baselines. We evaluate Average P@N across NeurIPS, SIGIR, and SciRepEval benchmarks, where ``Avg'' denotes the mean performance. P2R surpasses competitive baselines across diverse domains. Best results are in \textbf{bold}; second-best are \underline{underlined}.}
    \label{tab:main_results}
    
    \renewcommand{\arraystretch}{1.2}
    \setlength{\tabcolsep}{3pt} 
    
    \resizebox{\textwidth}{!}{
    \begin{tabular}{l ccc ccccc ccccc}
    \toprule

     & \multicolumn{3}{c}{\textbf{NeurIPS \cite{mimno2007expertise}}} 
     & \multicolumn{5}{c}{\textbf{SIGIR \cite{karimzadehgan2008multi}}} 
     & \multicolumn{5}{c}{\textbf{SciRepEval \cite{singh2023scirepeval}}} \\ 
    \cmidrule(lr){2-4} \cmidrule(lr){5-9} \cmidrule(lr){10-14}
    
    \textbf{Method} 
    & \makecell{\textbf{Soft} \\ \textbf{P@5}} & \makecell{\textbf{Hard} \\ \textbf{P@5}} & \cellcolor{gray!15}\textbf{AVG}
    & \textbf{Soft P@5} & \textbf{Soft P@10} & \textbf{Hard P@5} & \textbf{Hard P@10} & \cellcolor{gray!15}\textbf{AVG} 
    & \textbf{Soft P@5} & \textbf{Soft P@10} & \textbf{Hard P@5} & \textbf{Hard P@10} & \cellcolor{gray!15}\textbf{AVG} \\
    
    \midrule
    
    TPMS \cite{charlin2013toronto} 
        & 49.41 & 22.94 & \cellcolor{gray!15}36.18
        & 39.73 & 38.36 & 17.81 & 17.12 & \cellcolor{gray!15}28.26 
        & 62.06 & 53.74 & 31.40 & 24.86 & \cellcolor{gray!15}43.02 \\ 
        
    SciBERT \cite{beltagy2019scibert} 
        & 47.06 & 21.18 & \cellcolor{gray!15}34.12
        & 34.79 & 34.79 & 14.79 & 15.34 & \cellcolor{gray!15}24.93 
        & 59.63 & 54.39 & 28.04 & 24.49 & \cellcolor{gray!15}41.64 \\
        
    SPECTER \cite{cohan2020specter} 
        & 52.94 & 25.29 & \cellcolor{gray!15}39.12
        & 39.73 & 40.00 & 16.44 & 16.71 & \cellcolor{gray!15}28.22 
        & 65.23 & \textbf{56.07} & 32.34 & 25.42 & \cellcolor{gray!15}44.77 \\
        
    SciNCL \cite{ostendorff2022neighborhood} 
        & 54.12 & 27.06 & \cellcolor{gray!15}40.59
        & 40.55 & 39.45 & 17.81 & 17.40 & \cellcolor{gray!15}28.80 
        & 66.92 & 55.42 & 34.02 & 25.33 & \cellcolor{gray!15}45.42 \\
        
    COCO-DR \cite{yu2022coco} 
        & 54.12 & 25.29 & \cellcolor{gray!15}39.71
        & 40.00 & 40.55 & 16.71 & 17.53 & \cellcolor{gray!15}28.70 
        & 65.05 & 55.14 & 31.78 & 24.67 & \cellcolor{gray!15}44.16 \\
        
    SPECTER 2.0 CLF \cite{singh2023scirepeval} 
        & 52.35 & 24.71 & \cellcolor{gray!15}38.53
        & 39.45 & 38.63 & 16.16 & 16.30 & \cellcolor{gray!15}27.64 
        & 64.49 & 55.23 & 31.59 & 24.49 & \cellcolor{gray!15}43.95 \\
        
    SPECTER 2.0 PRX \cite{singh2023scirepeval} 
        & 54.12 & 27.65 & \cellcolor{gray!15}40.89
        & 37.53 & 37.12 & 16.44 & 16.05 & \cellcolor{gray!15}26.64 
        & 66.17 & 55.70 & 33.83 & \underline{25.61} & \cellcolor{gray!15}45.33 \\
        
    CoF \cite{zhang2025chain} 
        & \underline{55.68} & \underline{28.24} & \cellcolor{gray!15}\underline{41.96}
        & \underline{45.57} & \underline{41.69} & \textbf{22.47} & \underline{17.76} & \cellcolor{gray!15}\underline{31.87} 
        & \textbf{68.47} & 55.89 & \underline{34.52} & 25.33 & \cellcolor{gray!15}\underline{46.05} \\
    
    \midrule
    
    \textbf{P2R (Ours)} 
        & \textbf{59.41} & \textbf{33.38} & \cellcolor{gray!15}\textbf{46.40}
        & \textbf{46.85} & \textbf{43.29} & \underline{21.10} & \textbf{18.22} & \cellcolor{gray!15}\textbf{32.37} 
        & \underline{68.41} & \underline{55.98} & \textbf{35.05} & \textbf{27.58} & \cellcolor{gray!15}\textbf{46.76} \\
        
    \bottomrule
    \end{tabular}
    }
\end{table*}

\noindent\textbf{Evaluation Metrics. }
To quantitatively evaluate our model, we follow prior work\cite{singh2023scirepeval}. We adopt Precision@$N$ (P@$N$) as the primary evaluation metric, specifically reporting P@5 and P@10. 
For each submission paper $p$, let $\mathcal{R}_p$ denote the set of candidate reviewers for whom ground-truth relevance annotations exist. We rank all reviewers $r \in \mathcal{R}_p$ based on the relevance scores predicted by our model. Let $\hat{r}_{p, n}$ denote the reviewer ranked at the $n$-th position in this sorted list. The P@$N$ scores are then calculated under two settings: \textbf{Soft} and \textbf{Hard}, defined as follows:

\begin{equation} \label{eq:metrics}
\begin{aligned}
\text{Soft P}@N &= \frac{1}{|\mathcal{P}|} \sum_{p \in \mathcal{P}} \frac{\sum_{n=1}^N \mathbb{I}\left( y_{p, \hat{r}_{p, n}} \ge 2 \right)}{N}, \\
\text{Hard P}@N &= \frac{1}{|\mathcal{P}|} \sum_{p \in \mathcal{P}} \frac{\sum_{n=1}^N \mathbb{I}\left( y_{p, \hat{r}_{p, n}} = 3 \right)}{N}.
\end{aligned}
\end{equation}

Here, $|\mathcal{P}|$ denotes the total number of test papers, and $\mathbb{I}(\cdot)$ represents the indicator function. The term $y_{p, \hat{r}_{p, n}}$ denotes the ground-truth relevance score (ranging from 0 to 3) of the reviewer ranked at the $n$-th position for paper $p$.
Notably, due to the limited number of ground-truth reviewers per submission in NeurIPS (often fewer than ten), we restrict our evaluation to P@5 for this benchmark. For readability, we report P@N as percentages.

\noindent\textbf{Hyperparameter Settings. }
For PLM baselines, we follow~\cite{singh2023scirepeval} and aggregate paper--paper relevance into paper--reviewer relevance by averaging the top-3 matched papers.
For P2R, we set the RRF constant $k=60$, Top-$m$ aggregation $m=3$, and candidate size $M=15$.
We set the committee rubric weights $(\alpha,\beta,\gamma)=(0.5,0.3,0.2)$.
The discretization thresholds are $\mathcal{T}=\{\tau_1,\tau_2,\tau_3\}=\{0.35,1.35,2.35\}$ .
For RRF, we constrain weights to $w_i\in[0,2]$.

\subsection{Main results}

Table~\ref{tab:main_results} summarizes the comparative performance across all benchmarks. P2R establishes a new state-of-the-art, securing the top rank in 10 out of 13 metrics and ranking second in the remaining three.
This superiority is consistent across diverse domains, including machine learning (NeurIPS), information retrieval (SIGIR), and the broad scientific landscape (SciRepEval).
Notably, on the NeurIPS dataset, P2R achieves a substantial gain of +5.14 percentage points in Hard P@5 compared to the strongest baseline.
Since Hard P@5 strictly measures the retrieval of "very relevant" experts, this gain indicates that explicit aspect profiling captures fine-grained expertise nuances that implicit embeddings often miss.
Crucially, P2R achieves these results in a training-free manner, outperforming CoF which relies on task-specific fine-tuning. This suggests that leveraging LLMs for structured profiling and committee-based evaluation serves as a more effective paradigm than relying solely on latent representation learning.

\begin{table}[t]
    \centering

    \caption{Ablation analysis. Values are means of Soft and Hard P@N. Bold and \underline{underline} denote the best and second-best results.}
    \label{tab:ablation}
    
    \renewcommand{\arraystretch}{1.2} 
    \setlength{\tabcolsep}{4pt}
    
    \resizebox{\columnwidth}{!}{
    \begin{tabular}{lccc}
    \toprule
    \textbf{Model Variant} & \textbf{NeurIPS} & \textbf{SIGIR} & \textbf{SciRepEval} \\
    \midrule
    
    \multicolumn{4}{l}{\textit{\textbf{Individual Components}}} \\
    \hspace{3mm} Discrete Profile Matcher Only & 37.28 & 32.02 & 43.26 \\
    \hspace{3mm} Semantic Matcher Only & 45.32 & 26.79 & 46.61 \\
    \hspace{3mm} LLM Committee Only & 39.05 & 31.13 & 43.53 \\
    
    \midrule
    
    \multicolumn{4}{l}{\textit{\textbf{Hybrid Retrieval (Profile + Semantic)}}} \\
    \hspace{3mm} Hybrid Strategy (Discrete + Semantic) & \underline{45.62} & \underline{32.06} & \underline{46.64} \\
    
    \midrule
    
    \multicolumn{4}{l}{\textit{\textbf{Final Fusion (Hybrid + Committee)}}} \\
    \hspace{3mm} \textbf{P2R (Hybrid + LLM Committee)} & \textbf{46.40} & \textbf{32.37} & \textbf{46.76} \\
    
    \bottomrule
    \end{tabular}
    }
\end{table}
\subsection{Ablation experiment}
Our framework introduces two primary technical innovations. First, we implement LLM-driven structured profiling. We summarize schema-consistent expertise profiles from publication histories, explicitly spanning topics, methodologies, and applications. Second, we propose rubric-based score fusion. This module employs an LLM judge to assign aspect-level scores, which we integrate with hybrid retrieval signals via Reciprocal Rank Fusion (RRF). We validate the contribution of these components through a comprehensive ablation study. Specifically, we examine the following variants:

We compare five variants. \textbf{Discrete Profile Matcher Only} ranks reviewers by lexical overlap between flattened structured profiles. \textbf{Semantic Matcher Only} ranks reviewers by cosine similarity of dense embeddings. \textbf{LLM Committee Only} skips retrieval and directly applies the LLM committee to the full reviewer pool, testing whether LLMs can serve as standalone rankers without pre-filtering. \textbf{Hybrid Retrieval} combines the Discrete and Semantic matchers with RRF to produce a candidate list, without the final committee fusing. \textbf{P2R (Full Framework)} uses Hybrid Retrieval for candidate generation and then applies the rubric-driven committee for precision ranking, forming the complete coarse-to-fine pipeline.

Based on Table \ref{tab:ablation}, we highlight three key observations:
\textbf{Necessity of Pre-filtering.}
The standalone LLM Committee significantly underperforms the full framework. Direct evaluation over the global reviewer pool proves ineffective. The vast search space introduces excessive noise, distracting the model from granular expertise signals. A pre-filtered, high-quality candidate pool is indispensable for precise evaluation.
\textbf{Benefits of Hybrid Retrieval.}
Integrating discrete and continuous signals yields consistent gains.
The Hybrid Strategy outperforms both single-stream baselines.
This confirms that the structural precision of profiles complements the semantic coverage of embeddings.
\textbf{Effectiveness of Coarse-to-Fine Strategy.}
The full P2R model achieves optimal performance.
Retrieval provides a high-quality candidate pool, and subsequent committee evaluation enhances precision.
This validates our design choices.
Combining hybrid retrieval for candidate generation and the LLM committee for final refinement yields the best performance.

These results confirm that each P2R component is indispensable.
The framework effectively enhances traditional retrieval with LLM-based evaluating, leveraging their distinct strengths for accurate reviewer matching.

\begin{table}[t]
    \centering
\caption{Performance comparison of profile representations. Values represent the average of soft and hard P@N metrics. Best results are in \textbf{bold}}
    \label{tab:representation_comparison}
    \renewcommand{\arraystretch}{1.15}
    \resizebox{\columnwidth}{!}{%
        \setlength{\tabcolsep}{8pt}
        \begin{tabular}{lccc}
            \toprule
            \textbf{Representation Strategy} & \textbf{NeurIPS} & \textbf{SIGIR} & \textbf{SciRepEval} \\
            \midrule

            Unstructured Description & 45.81 &25.79 & 46.52 \\
            \textbf{Structured Aspects(Ours)} & \textbf{46.40} & \textbf{32.37} & \textbf{46.76} \\

            \bottomrule
        \end{tabular}%
    }
\end{table}

\begin{table}[!t]
    \centering
    \caption{Impact of profile granularity. Values represent the average of soft and hard P@N metrics. Best results are in \textbf{bold}; second-best are \underline{underlined}.}
    \label{tab:profile_ablation}

    \footnotesize
    \renewcommand{\arraystretch}{1.1}
    \setlength{\tabcolsep}{6pt}

    \begin{tabular}{lccc}
    \toprule
    \textbf{Configuration} & \textbf{NeurIPS} & \textbf{SIGIR} & \textbf{SciRepEval} \\
    \midrule
    Best Subset ($j=1$) & \underline{46.10} & 28.90 & 46.36 \\
    Best Subset ($j=2$) & 45.52 & \underline{30.82} & \underline{46.68} \\
    \textbf{P2R (Full Profile)} & \textbf{46.40} & \textbf{32.37} & \textbf{46.76} \\
    \bottomrule
    \end{tabular}
\end{table}

\subsection{Impact of Profile Design} 

We investigate the optimal construction of expertise profiles from two perspectives: representation format and aspect granularity. \noindent\textbf{Structured vs. Unstructured Representation.} Table~\ref{tab:representation_comparison} substantiates the superiority of \textit{Structured Aspects} over \textit{Unstructured Descriptions} (free-form summaries). While free-form summaries capture broader context, they are prone to introducing generic noise and hallucinations, particularly when reviewer histories are sparse. In contrast, structured profiling acts as a semantic filter. By enforcing a strict schema, the model is constrained to discard irrelevant information and extract precise technical keywords. This "denoising" effect explains the consistent performance gains across all datasets, validating structured profiling as a more robust representation for expertise matching.
\noindent\textbf{Aspect Granularity.}
We further analyze the contribution of different expertise dimensions. Table~\ref{tab:profile_ablation} reports the peak performance for aspect subsets of size $j$.
A clear trend emerges: single-aspect matching is insufficient for accurate recommendation.
The complete triad—\textit{Topics}, \textit{Methodologies}, and \textit{Applications}—consistently yields the best performance.
These results confirm that expertise matching requires coverage across multiple dimensions. \textit{Topics} align the broad subject, \textit{Methodologies} verify technical fit, and \textit{Applications} define the target setting. These signals are additive. Dropping any single dimension removes critical context, causing the model to miss subtle mismatches.

\subsection{Committee Configuration Analysis}
How to configure the committee critically impacts evaluation accuracy. Here, we examine these two fundamental design choices.
\noindent\textbf{Ranking vs. Scoring.}
Table~\ref{tab:scoring_strategy} shows that Rubric Scoring outperforms Direct Ranking across all datasets. 
This difference is primarily due to the cognitive load imposed by each method. Direct Ranking requires comparing multiple candidates at once, often leading to vague or imprecise judgments. In contrast, Rubric Scoring breaks the decision into individual checks, prompting the model to assess each candidate against specific criteria (e.g., "Does the methodology match?"). This approach ensures that decisions are based on clear, evidence-backed evaluation, reducing ambiguity and aligning better with human review standards.
\noindent\textbf{Committee Composition.}
Table~\ref{tab:ablation_agents} confirms the necessity of a diverse committee.
The full configuration consistently surpasses smaller subsets, indicating that distinct roles capture more sufficient signals for robust evaluation.
For instance, a "Theorist" expert identifies mathematical mismatches that a generalist might overlook, while an "Application" expert ensures domain fit.
A single perspective leaves information gaps, whereas the full committee provides a more complete and robust evaluation.
\subsection{Model Robustness}

\begin{table}[!t]
    \centering
    \caption{Impact of Decision Strategy. Values represent the average of soft and hard P@N metrics.}
    \label{tab:scoring_strategy}
    \begin{tabular}{l|ccc}
        \toprule
        \textbf{Strategy} & \textbf{NeurIPS} & \textbf{SIGIR} & \textbf{SciRepEval} \\
        \midrule
        Direct Ranking & 42.58 & 31.59 & 46.27 \\
        \textbf{Rubric Scoring (Ours)} & \textbf{46.40} & \textbf{32.37} & \textbf{46.76} \\
        \bottomrule
    \end{tabular}
\end{table}

\begin{table}[t]
    \centering
    \footnotesize 
    \caption{Impact of committee size. Values represent the average of soft and hard P@N metrics. Best results are in \textbf{bold}}
    \label{tab:ablation_agents}
    
    \renewcommand{\arraystretch}{1.0} 
    \setlength{\tabcolsep}{6pt}     
    
    \begin{tabular}{lccc}
    \toprule
    \textbf{Configuration} & \textbf{NeurIPS} & \textbf{SIGIR} & \textbf{SciRepEval} \\
    \midrule
    Best Subset ($C=1$) & 45.80 & 32.12 & 46.54 \\
    Best Subset ($C=2$) & \underline{46.10} & 32.10 & 46.47 \\
    Best Subset ($C=3$) & 45.81 & \underline{32.13} & \underline{46.57} \\
    \midrule
    \textbf{P2R (Full, $C=4$)} & \textbf{46.40} & \textbf{32.37} & \textbf{46.76} \\
    \bottomrule
    \end{tabular}
\end{table}

\begin{figure}[!t]
    \centering
        \subfloat[]{
        \includegraphics[height=3.2cm]{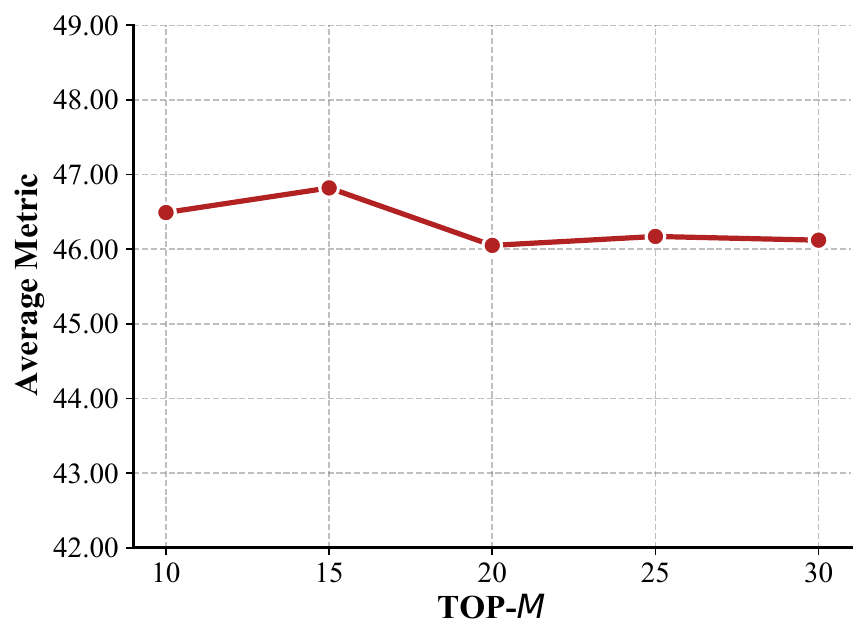}
        \label{fig:sensitivity_topm}
    }
    \hfill
    \subfloat[]{
        \includegraphics[height=3.2cm]{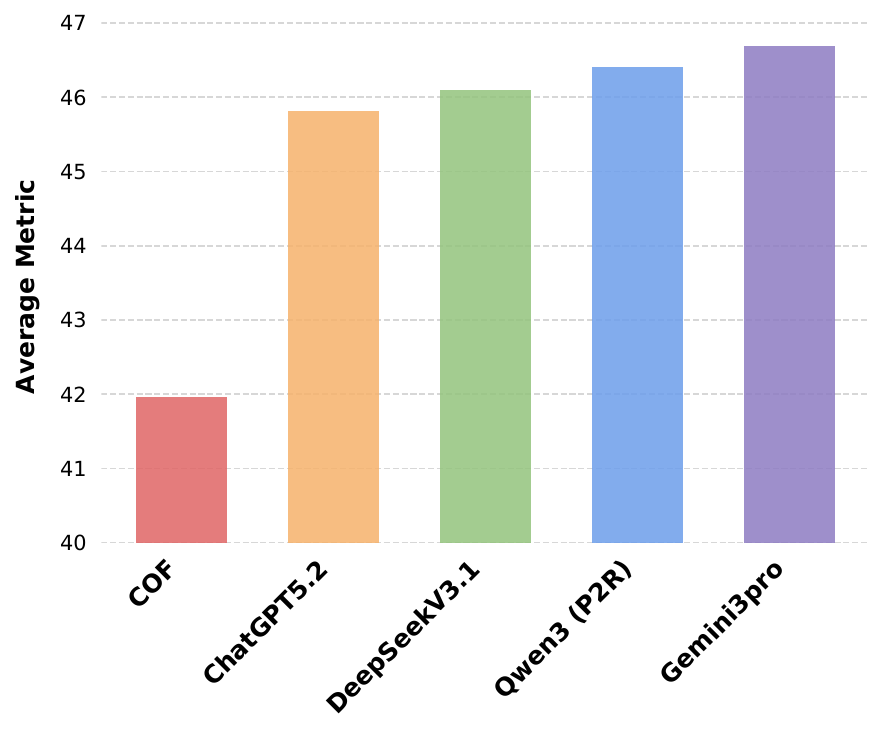}
        \label{fig:robust_llm}
    }
\caption{(a) Top-$M$ sensitivity on SciRepEval. (b) LLM-backbone robustness on NeurIPS. Values represent the average of soft and hard P@N metrics.}

    \label{fig:robust_and_topm}
\end{figure}

We further examine the stability of P2R from two perspectives: sensitivity to the candidate pool size and robustness to different LLM backbones.
After the hybrid retrieval stage, we retain the Top-$M$ reviewers as the candidate set $\mathcal{C}_p$ for the fine-grained committee evaluation.
As shown in Fig.~\ref{fig:sensitivity_topm}, varying $M$ leads to only minor fluctuations in the overall performance on SciRepEval, indicating that P2R is not sensitive to the choice of $M$.
To assess the generalization capability of P2R, we conduct studies using various LLM backbones.
We instantiate P2R with diverse backbones, including ChatGPT5.2, DeepSeekV3.1, Qwen3, and Gemini3Pro, and evaluate them on the NeurIPS dataset (Fig.~\ref{fig:robust_llm}).
We benchmark these variants against Chain-of-Factors (CoF)\cite{zhang2025chain}, the state-of-the-art baseline.
Regardless of the underlying LLM backbone, P2R consistently surpasses CoF, a scientific document representation model fine-tuned for reviewer matching.
Notably, various general-purpose models achieve strong results, confirming the framework's robustness.
These results confirm that our performance gains derive from the structured profiling mechanism rather than model-specific capabilities.

\section{Conclusion}
In this paper, we proposed P2R, a training-free framework that addresses the lack of explicit and multi-dimensional modeling of reviewers in prevalent approaches. P2R summarizes structured profiles spanning Topics, Methodologies, and Applications for both submissions and reviewers. It employs an LLM-based, rubric-driven committee to emulate the  multi-dimensional assessment of human Area Chairs. Experiments on NeurIPS, SIGIR, and SciRepEval verify the effectiveness of our framework. P2R consistently surpasses state-of-the-art baselines across these datasets. Comprehensive ablation studies validate that each module within our framework is effective and indispensable. Finally, this work not only provides an effective solution to the reviewer matching challenge but also offers valuable empirical insights for the application of LLMs in this specific domain.

\bibliography{rf}

@article{price2017computational,
  title={Computational support for academic peer review: a perspective from artificial intelligence},
  author={Price, Simon and Flach, Peter A},
  journal={Communications of the ACM},
  volume={60},
  number={3},
  pages={70--79},
  year={2017},
  publisher={ACM New York, NY, USA}
}

@inproceedings{mimno2007expertise,
  title={Expertise modeling for matching papers with reviewers},
  author={Mimno, David and McCallum, Andrew},
  booktitle={Proceedings of the 13th ACM SIGKDD international conference on Knowledge discovery and data mining},
  pages={500--509},
  year={2007}
}

@inproceedings{charlin2011framework,
  title={A Framework for Optimizing Paper Matching.},
  author={Charlin, Laurent and Zemel, Richard S and Boutilier, Craig},
  booktitle={UAI},
  volume={11},
  pages={86--95},
  year={2011}
}

@inproceedings{charlin2013toronto,
  title     = {The Toronto Paper Matching System: An Automated Paper-Reviewer Assignment System},
  author    = {Charlin, Laurent and Zemel, Richard S.},
  booktitle = {ICML 2013 Workshop on Peer Reviewing and Publishing Models},
  year      = {2013}
}

@inproceedings{liu2014robust,
  title={A robust model for paper reviewer assignment},
  author={Liu, Xiang and Suel, Torsten and Memon, Nasir},
  booktitle={Proceedings of the 8th ACM Conference on Recommender systems},
  pages={25--32},
  year={2014}
}

@inproceedings{zhang2025chain,
  title={Chain-of-factors paper-reviewer matching},
  author={Zhang, Yu and Shen, Yanzhen and Kang, SeongKu and Chen, Xiusi and Jin, Bowen and Han, Jiawei},
  booktitle={Proceedings of the ACM on Web Conference 2025},
  pages={1901--1910},
  year={2025}
}

@inproceedings{tang2010expertise,
  title={Expertise matching via constraint-based optimization},
  author={Tang, Wenbin and Tang, Jie and Tan, Chenhao},
  booktitle={2010 IEEE/WIC/aCM international conference on web intelligence and intelligent agent technology},
  volume={1},
  pages={34--41},
  year={2010},
  organization={IEEE}
}

@article{yang2025qwen3,
  title={Qwen3 technical report},
  author={Yang, An and Li, Anfeng and Yang, Baosong and Zhang, Beichen and Hui, Binyuan and Zheng, Bo and Yu, Bowen and Gao, Chang and Huang, Chengen and Lv, Chenxu and others},
  journal={arXiv preprint arXiv:2505.09388},
  year={2025}
}

@article{peng2025frontier,
  title={FRONTIER-RevRec: A Large-scale Dataset for Reviewer Recommendation},
  author={Peng, Qiyao and Wang, Chen and Wang, Yinghui and Liu, Hongtao and Guo, Xuan and Wang, Wenjun},
  journal={arXiv preprint arXiv:2510.16597},
  year={2025}
}

@inproceedings{tong2006fast,
  title={Fast random walk with restart and its applications},
  author={Tong, Hanghang and Faloutsos, Christos and Pan, Jia-Yu},
  booktitle={Sixth international conference on data mining (ICDM'06)},
  pages={613--622},
  year={2006},
  organization={IEEE}
}

@inproceedings{beltagy2019scibert,
  title     = {SciBERT: A Pretrained Language Model for Scientific Text},
  author    = {Beltagy, Iz and Lo, Kyle and Cohan, Arman},
  booktitle = {Proceedings of the 2019 Conference on Empirical Methods in Natural Language Processing and the 9th International Joint Conference on Natural Language Processing (EMNLP-IJCNLP)},
  pages     = {3615--3620},
  year      = {2019},
  doi       = {10.18653/v1/D19-1371}
}

@inproceedings{cohan2020specter,
  title     = {SPECTER: Document-Level Representation Learning using Citation-Informed Transformers},
  author    = {Cohan, Arman and Feldman, Sergey and Beltagy, Iz and Downey, Doug and Weld, Daniel},
  booktitle = {Proceedings of the 58th Annual Meeting of the Association for Computational Linguistics},
  pages     = {2270--2282},
  year      = {2020},
  doi       = {10.18653/v1/2020.acl-main.207}
}

@inproceedings{ostendorff2022neighborhood,
  title     = {Neighborhood Contrastive Learning for Scientific Document Representations with Citation Embeddings},
  author    = {Ostendorff, Malte and Rethmeier, Nils and Augenstein, Isabelle and Gipp, Bela and Rehm, Georg},
  booktitle = {Proceedings of the 2022 Conference on Empirical Methods in Natural Language Processing},
  pages     = {11670--11688},
  year      = {2022},
  doi       = {10.18653/v1/2022.emnlp-main.802}
}

@inproceedings{singh2023scirepeval,
  title={Scirepeval: A multi-format benchmark for scientific document representations},
  author={Singh, Amanpreet and D’Arcy, Mike and Cohan, Arman and Downey, Doug and Feldman, Sergey},
  booktitle={Proceedings of the 2023 Conference on Empirical Methods in Natural Language Processing},
  pages={5548--5566},
  year={2023}
}

@inproceedings{cormack2009reciprocal,
  title={Reciprocal rank fusion outperforms condorcet and individual rank learning methods},
  author={Cormack, Gordon V and Clarke, Charles LA and Buettcher, Stefan},
  booktitle={Proceedings of the 32nd international ACM SIGIR conference on Research and development in information retrieval},
  pages={758--759},
  year={2009}
}

@inproceedings{yu2022coco,
  title     = {COCO-DR: Combating the Distribution Shift in Zero-Shot Dense Retrieval with Contrastive and Distributionally Robust Learning},
  author    = {Yu, Yue and Xiong, Chenyan and Sun, Si and Zhang, Chao and Overwijk, Arnold},
  booktitle = {Proceedings of the 2022 Conference on Empirical Methods in Natural Language Processing},
  pages     = {1462--1479},
  year      = {2022},
  doi       = {10.18653/v1/2022.emnlp-main.95}
}

@article{wang2026ai,
  author  = {Jialiang Wang and Yuchen Liu and Hang Xu and Kaichun Hu and
             Shimin Di and Wangze Ni and Linan Yue and Min-Ling Zhang and
             Kui Ren and Lei Chen},
  title   = {When {AI} Reviews Science: Can We Trust the Referee?},
  journal = {The Innovation Informatics},
  volume  = {2},
  number  = {1},
  pages   = {100030},
  year    = {2026},
  doi     = {10.59717/j.xinn-inform.2026.100030}
}

@inproceedings{lee2024nv,
  title     = {NV-Embed: Improved Techniques for Training LLMs as Generalist Embedding Models},
  author    = {Lee, Chankyu and Roy, Rajarshi and Xu, Mengyao and Raiman, Jonathan and Shoeybi, Mohammad and Catanzaro, Bryan and Ping, Wei},
  booktitle = {International Conference on Learning Representations},
  year      = {2025}
}

@book{schrijver2003combinatorial,
  title={Combinatorial optimization: polyhedra and efficiency},
  author={Schrijver, Alexander and others},
  volume={24},
  number={2},
  year={2003},
  publisher={Springer}
}

@inproceedings{sun2023chatgpt,
  title     = {Is ChatGPT Good at Search? Investigating Large Language Models as Re-Ranking Agents},
  author    = {Sun, Weiwei and Yan, Lingyong and Ma, Xinyu and Wang, Shuaiqiang and Ren, Pengjie and Chen, Zhumin and Yin, Dawei and Ren, Zhaochun},
  booktitle = {Proceedings of the 2023 Conference on Empirical Methods in Natural Language Processing},
  pages     = {14918--14937},
  year      = {2023},
  doi       = {10.18653/v1/2023.emnlp-main.923}
}

@inproceedings{hou2024large,
  title={Large language models are zero-shot rankers for recommender systems},
  author={Hou, Yupeng and Zhang, Junjie and Lin, Zihan and Lu, Hongyu and Xie, Ruobing and McAuley, Julian and Zhao, Wayne Xin},
  booktitle={European Conference on Information Retrieval},
  pages={364--381},
  year={2024},
  organization={Springer}
}

@article{zheng2023judging,
  title={Judging llm-as-a-judge with mt-bench and chatbot arena},
  author={Zheng, Lianmin and Chiang, Wei-Lin and Sheng, Ying and Zhuang, Siyuan and Wu, Zhanghao and Zhuang, Yonghao and Lin, Zi and Li, Zhuohan and Li, Dacheng and Xing, Eric and others},
  journal={Advances in neural information processing systems},
  volume={36},
  pages={46595--46623},
  year={2023}
}

@article{gu2024survey,
  title   = {A Survey on LLM-as-a-Judge},
  author  = {Gu, Jiawei and Jiang, Xuhui and Shi, Zhichao and Tan, Hexiang and Zhai, Xuehao and Xu, Chengjin and Li, Wei and Shen, Yinghan and Ma, Shengjie and Liu, Honghao and others},
  journal = {The Innovation},
  pages   = {101253},
  year    = {2026},
  doi     = {10.1016/j.xinn.2025.101253}
}

@article{saveski2023counterfactual,
  title={Counterfactual evaluation of peer-review assignment policies},
  author={Saveski, Martin and Jecmen, Steven and Shah, Nihar and Ugander, Johan},
  journal={Advances in Neural Information Processing Systems},
  volume={36},
  pages={58765--58786},
  year={2023}
}

@inproceedings{karan2025dataset,
  title={A Dataset for Expert Reviewer Recommendation with Large Language Models as Zero-shot Rankers},
  author={Karan, Vanja M and McQuistin, Stephen and Yanagida, Ryo and Perkins, Colin and Tyson, Gareth and Castro, Ignacio and Healey, Patrick and Purver, Matthew},
  booktitle={Proceedings of the 31st International Conference on Computational Linguistics},
  pages={11422--11427},
  year={2025}
}

@inproceedings{karimzadehgan2008multi,
  title={Multi-aspect expertise matching for review assignment},
  author={Karimzadehgan, Maryam and Zhai, ChengXiang and Belford, Geneva},
  booktitle={Proceedings of the 17th ACM conference on Information and knowledge management},
  pages={1113--1122},
  year={2008}
}

@article{ma2023comprehensive,
  title={A comprehensive survey on vector database: Storage and retrieval technique, challenge},
  author={Ma, Le and Zhang, Ran and Han, Yikun and Yu, Shirui and Wang, Zaitian and Ning, Zhiyuan and Zhang, Jinghan and Xu, Ping and Li, Pengjiang and Ju, Wei and others},
  journal={arXiv preprint arXiv:2310.11703},
  year={2023}
}

@inproceedings{cai2023resolving,
  title={Resolving the imbalance issue in hierarchical disciplinary topic inference via llm-based data augmentation},
  author={Cai, Xunxin and Xiao, Meng and Ning, Zhiyuan and Zhou, Yuanchun},
  booktitle={2023 IEEE international conference on data mining workshops (ICDMW)},
  pages={1424--1429},
  year={2023},
  organization={IEEE}
}

@article{dong2023adaptive,
  title={Adaptive path-memory network for temporal knowledge graph reasoning},
  author={Dong, Hao and Ning, Zhiyuan and Wang, Pengyang and Qiao, Ziyue and Wang, Pengfei and Zhou, Yuanchun and Fu, Yanjie},
  journal={arXiv preprint arXiv:2304.12604},
  year={2023}
}

@article{ning2022graph,
  title={Graph soft-contrastive learning via neighborhood ranking},
  author={Ning, Zhiyuan and Wang, Pengfei and Wang, Pengyang and Qiao, Ziyue and Fan, Wei and Zhang, Denghui and Du, Yi and Zhou, Yuanchun},
  journal={arXiv preprint arXiv:2209.13964},
  year={2022}
}

@inproceedings{ning2021lightcake,
  title={Lightcake: A lightweight framework for context-aware knowledge graph embedding},
  author={Ning, Zhiyuan and Qiao, Ziyue and Dong, Hao and Du, Yi and Zhou, Yuanchun},
  booktitle={Pacific-Asia Conference on Knowledge Discovery and Data Mining},
  pages={181--193},
  year={2021},
  organization={Springer}
}

@article{qiao2022rpt,
  title={Rpt: toward transferable model on heterogeneous researcher data via pre-training},
  author={Qiao, Ziyue and Fu, Yanjie and Wang, Pengyang and Xiao, Meng and Ning, Zhiyuan and Zhang, Denghui and Du, Yi and Zhou, Yuanchun},
  journal={IEEE Transactions on Big Data},
  volume={9},
  number={1},
  pages={186--199},
  year={2022},
  publisher={IEEE}
}

@article{xiao2025interdisciplinary,
  title={Interdisciplinary fairness in imbalanced research proposal topic inference: A hierarchical transformer-based method with selective interpolation},
  author={Xiao, Meng and Wu, Min and Qiao, Ziyue and Fu, Yanjie and Ning, Zhiyuan and Du, Yi and Zhou, Yuanchun},
  journal={ACM Transactions on Knowledge Discovery from Data},
  volume={19},
  number={2},
  pages={1--21},
  year={2025},
  publisher={ACM New York, NY}
}

@article{qiao2020context,
  title={Context-enhanced entity and relation embedding for knowledge graph completion},
  author={Qiao, Ziyue and Ning, Zhiyuan and Du, Yi and Zhou, Yuanchun},
  journal={arXiv preprint arXiv:2012.07011},
  year={2020}
}

@article{ning2025deep,
  title={Deep cut-informed graph embedding and clustering},
  author={Ning, Zhiyuan and Wang, Zaitian and Zhang, Ran and Xu, Ping and Liu, Kunpeng and Wang, Pengyang and Ju, Wei and Wang, Pengfei and Zhou, Yuanchun and Cambria, Erik and others},
  journal={Information Fusion},
  pages={103603},
  year={2025},
  publisher={Elsevier}
}

@article{ning2025rethinking,
  title={Rethinking graph contrastive learning through relative similarity preservation},
  author={Ning, Zhiyuan and Wang, Pengfei and Qiao, Ziyue and Wang, Pengyang and Zhou, Yuanchun},
  journal={arXiv preprint arXiv:2505.05533},
  year={2025}
}

@article{weijun2023identifying,
  title={Identifying Interdisciplinary Sci-Tech Literature Based on Multi-Label Classification},
  author={Weijun, Wang and Zhiyuan, Ning and Du, Y and Yuanchun, Z},
  journal={Data Analysis and Knowledge Discovery},
  volume={7},
  number={1},
  pages={102--112},
  year={2023}
}

@article{ning2024fedgcs,
  title={FedGCS: A generative framework for efficient client selection in federated learning via gradient-based optimization},
  author={Ning, Zhiyuan and Tian, Chunlin and Xiao, Meng and Fan, Wei and Wang, Pengyang and Li, Li and Wang, Pengfei and Zhou, Yuanchun},
  journal={arXiv preprint arXiv:2405.06312},
  year={2024}
}
\bibliographystyle{IEEEtran}

\end{document}